# A modified Moss rule highlights underexplored classes of high refractive index materials

Eugène Bertin*, Finja Tadge, Andrea Crovetto*

*National Centre for Nano Fabrication and Characterization (DTU Nanolab), Technical University of Denmark, Building 347 East, Ørsteds Pl., 2800 Lundtofte, Denmark*

**corresponding authors*

## Abstract

High refractive index dielectrics are central to photonic applications, yet the empirical Moss rule imposes a fundamental trade-off between refractive index and optical transparency. We introduce a modified Moss rule anchored to the optical absorption edge rather than the fundamental bandgap, capturing materials where various physical mechanisms suppress absorption well above the electronic gap. Screening the Materials Project database with this metric reveals recurring chemical compositions and structural prototypes in the materials with the most promising refractive index/transparency trade-offs. These materials include chalcopyrites, Zintl pnictides, and early transition metal multi-anion chalconitrides and oxichalcogenide compounds. Hybrid density functional theory calculations reveal that the chalconitride $(Hf,Zr)_2(S,Se)N_2$ family can achieve a combination of ultra-high refractive indices, low effective masses and transparency windows extending into the UV, potentially surpassing state-of-the-art $TiO_2$. Our results establish $(Hf,Zr)_2(S,Se)N_2$ compounds as a unique and largely unexplored material family for next-generation photonic applications.

## Introduction

High refractive index (n) dielectric materials are central to photonic technologies ranging from nanoresonators and metalenses to nonlinear optical devices, where all performance indicators scale with refractive index [1]. However, the empirical Moss rule states that achieving high index alongside low optical absorption is fundamentally difficult. This rule expresses the relationship between the sub-bandgap refractive index $n_0$ and the bandgap energy $E_g$ as

$$E_g n_0^4 = 95\ eV,$$

meaning that high transparency and large refractive index tend to be mutually exclusive. Identifying super-Mossian materials that surpass this rule has therefore become an ongoing effort in optical materials research [1,2].

In the standard dielectric-dispersion framework, $n_0$ is governed primarily by the Penn gap $E_P$, an average interband transition energy weighted by the joint density of states, rather than by the fundamental bandgap $E_g$ [3]. Provided that optical transparency for photon energies of interest is maintained, a high refractive index is therefore obtained when $E_P$ sits closer to those energies. This corresponds to a case where most of the spectral weight of absorption is clustered near and above the photon energy of interest. Therefore, it is of paramount importance to accurately consider the spectral distribution of absorption when screening for high index semiconductors.

Many semiconductors and dielectrics have forbidden or weak optical transitions near their fundamental bandgap, such that absorption remains negligible well above the electronic gap. Woods-Robinson et al. [4] showed that screening for optical transparency based on the fundamental gap alone systematically overlooks materials whose absorption edge is pushed to much higher photon energies by various physical mechanisms. These include symmetry and momentum selection rules, and negligible wavefunction overlap between the initial and final states. All these factors can lead to low transition dipole moments. This framework motivated our recent experimental demonstration of $Zr_2SN_2$ as thin films with a wide transparency range and large refractive index [5], even compared to state-of-the-art materials such as rutile $TiO_2$ (Table S2). $Zr_2SN_2$ would have been excluded from any conventional transparency screening funnel due to its indirect bandgap of only 1.43 eV calculated with hybrid density functional theory (DFT). Yet, weak indirect transitions up to its 2.63 eV direct gap, followed by a wide window of forbidden transitions, suppress absorption up until 3.4 eV photon energy.

Although suppressed absorption does not contribute to the sub-bandgap refractive index, forbidden or weak optical transitions may decrease the difference between the Penn gap $E_P$ and the characteristic energy where absorption becomes significant, such as the lowest direct allowed transition, enabling super-Mossian behaviour. In this work, we build on the role of forbidden transitions and propose a modified Moss rule anchored not to the fundamental gap $E_g$ but to the absorption edge energy $E_{edge}$ i.e. the energy at which absorption becomes non-negligible. We query the Materials Project [6] for the two key quantities required for assessing high refractive index materials' performance. The first is the absorption coefficient α spectra computed by Woods-Robinson et al. [4], defining $E_{edge}$ at the threshold of $\alpha = 10^4$ $cm^{-1}$. The second is the high-frequency dielectric constant $\varepsilon_\infty$ [7,8] used to obtain the sub-bandgap refractive index $n_0$ through $n_0 = \sqrt{\varepsilon_\infty}$. We define a modified Moss factor $M_{edge}$ as

$$M_{edge} = \frac{E_{edge} n_0^4}{95\ eV}\ ;$$

this single substitution of the fundamental gap $E_g$ by the absorption edge $E_{edge}$ shifts the figure of merit from a purely electronic quantity to one that reflects a material's actual onset of optical losses. Materials with $M_{edge} = 1$ behave as heuristically predicted by the Moss rule, while materials with $M_{edge} > 1$ perform better than predicted and can be qualified as super-Mossian [1].

## Methods

We query Materials Project for materials combining a modified Moss factor above 1.2 with a PBE absorption edge above 1.6 eV. Since the semi-local PBE functional [9] used throughout Materials Project systematically underestimates bandgaps, the true absorption edges of this set of high-n compounds are expected to fall close or above the visible/UV edge, with transparency potentially spanning the entire visible spectrum.

Thermodynamic stability of the screened materials is enforced by only considering materials that lie within 0.05 eV/atom of the convex stability hull. We validate the method based on binary compounds, and verify that materials recognized for their high refractive index like BP, GaP, AlAs, $TiO_2$, $SnS_2$ [1,10,11] are highlighted by our analysis (Figure S1). We constrain the rest of the study to candidate ternary and quaternary compounds. Frequency-dependent density functional theory (DFT) calculations employing the hybrid HSE06 functional [12,13] with 25 % Hartree-Fock exchange and a screening parameter of 0.208 $Å^{-1}$ were conducted to obtain refractive index and absorption coefficient spectra for the promising candidates of the $(Hf,Zr)_2(S,Se)N_2$ family. All calculations were performed with the independent particle approximation within the LOPTICS framework, as implemented in the Vienna Ab-Initio Simulation Package (VASP) [14–18] using a plane wave energy cutoff of 550 eV. We note that this method neglects other possible sources of absorption, including indirect transitions and excitonic effects. Both phenomena could lead to a decrease in the span of the transparent regions, and therefore in the Moss factor. Regular Γ-centered k-point meshes including $12 \times 12 \times 6$ and $12 \times 12 \times 3$ k-points were employed for the $\alpha$- and $\beta$-polymorphs defined in the article, respectively. To smooth the discrete interband transition energies arising from the finite k-point mesh, the calculated spectra were convoluted with a Gaussian broadening of 0.1 eV. This numerical broadening does not model the physical Urbach tail, which is typically characterized by an exponential energy dependence. Electron and hole effective masses in different crystal directions are obtained from non self-consistent HSE06 band structure calculations performed on a high symmetry path via parabolic fitting of the corresponding band edges using the sumo package [19].

## Results and Discussion

We report the modified Moss factor $M_{edge}$ and PBE absorption edge $E_{edge}$ for the resulting dataset of high-performing high-n low loss materials (Figure 1a). $BeSiP_2$, still experimentally unreported, was identified by Woods-Robinson et al. [4] as a potential transparent conductor owing to its exceptionally wide range of forbidden transitions. A similar observation can be made for ZrOS [20]. Materials already demonstrated or screened as high-n transparent materials appear, such as $Zr_2SN_2$ [4,5], validating the method. Interestingly, $BeSiP_2$, $Zr_2SN_2$ and ZrOS were all screened as candidate transparent conductors; this establishes forbidden transitions as a shared design handle for transparent conductors and high-n optical materials.

Compositionally diverse compounds stand out as potential super-Mossian materials (Figure 1a); a few notable experimentally reported examples are $ZnSiP_2$, $AlCuTe_2$, $Zr_2SN_2$, $Mg(BeAs)_2$, NaCuTe, $Mg(BeP)_2$, $Hf_2SN_2$, $CuBS_2$, ZrSO, $Al_2BiSe_4$ and HfSO. Several experimentally unrealized materials such as $BeSiAs_2$, $Hf_2SeN_2$, $BeSiP_2$ also stand out. Importantly, many of these compounds have forbidden transitions at the lowest direct bandgap (diamond-shaped markers on Figure 1a – as calculated by Woods-Robinson et al. [4]), contributing to their high $M_{edge}$.

Beyond uncovering individual candidate compounds, this dataset reveals interesting links between elemental composition, transparency and high refractive index, as theorized by Naccarato et al. [21] and Arai et al. [20]. We extract the most common elements in the dataset (Figure 1b); the marker size scales linearly with the number of compounds containing each element. Alkali and alkaline earth metals rank high in terms of average $M_{edge}$ and are overrepresented, particularly Be with 7 occurrences. However, Be's high toxicity makes it hazardous to handle, diminishing the technological relevance and ease of production of such compounds. Si, already established as a photonic platform in its elemental semiconductor form, is also present in several promising materials. Early- and post-transition metals are also well represented. In particular Zr and Hf early transition metals in the $d^0$ electronic configuration seem particularly promising, due to the conjunction of high $E_{edge}$ and $M_{edge}$, in accordance with Naccarato's et al. [21] chemical design.

Although O-containing compounds often exhibit high values of $E_{edge}$, there are many super-Mossian non-oxide chalcogenides, nitrides and phosphides, and they generally have higher values of the $M_{edge}$ figure of merit. Interestingly, all high performing O-containing compounds (except $Sc_2Pt_2O_7$) also contain a second anion, for example HfSO and TaNO. More generally, a significant fraction of super-Mossian compounds like $Hf_2SN_2$ belong to multi-anion chemistries. The above arguments hint at the importance of moving beyond single-anion oxide chemistry for the discovery of exceptional super-Mossian materials.

To identify recurring structure prototypes, we group compounds that share identical chemical formula template and space group (Figure 1c). Prototypes are labelled based on the highest-$M_{edge}$ compounds of the group, and compounds belonging to each group are coloured accordingly in Figure 1a. For example, compounds from the most represented group (8 representatives), labelled $BeSiAs_2$ (formula: $ABC_2$ | space group: I-42d), share the same zinc-blende-derived chalcopyrite structure. This family contains chalcogenides of Group IB and Group IIIA elements (Cu or Ag on the A site; Al or B on the B site) as well as pnictides with a Group IIA/IIB metal on the A site (Zn, Mg, Be), and Si or Ge on the B site. This fits previous observations of exceptional refractive index and exotic optical properties [22] in other chalcopyrite phosphides such as $ZnGeP_2$ [23,24]. The $Mg(BeAs)_2$ group ($AB_2C_2$ | P-3m1) hosts alkaline earth Zintl pnictides [25] ($Ca(MgAs)_2$, $Mg(BeAs)_2$ and $Mg(BeP)_2$), hinting at the super-Mossian potential of this class material, in accordance with ref. [25].

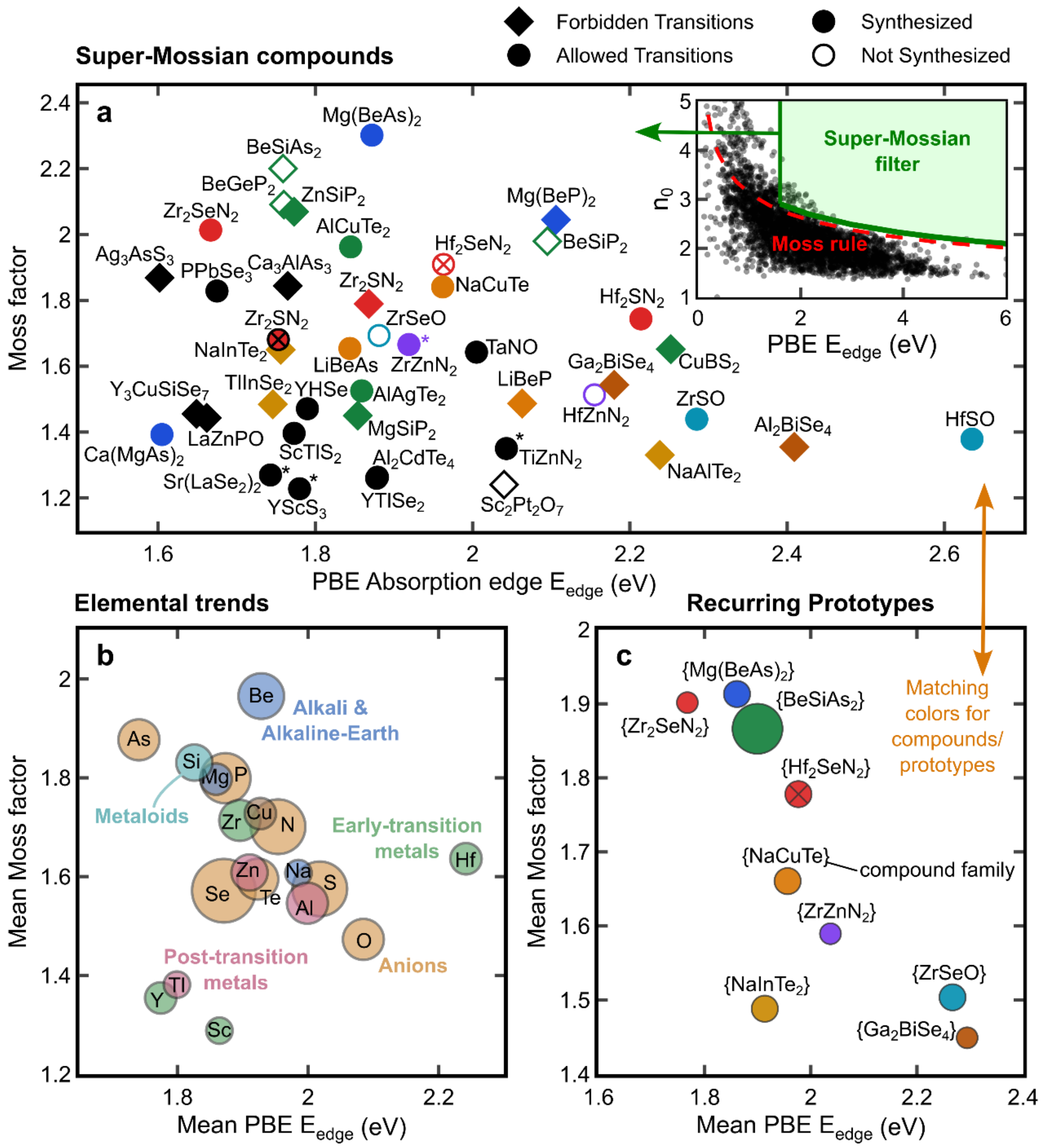

*Figure 1: Potential super-Mossian materials according to the calculated PBE-level optical properties available on Materials Project. (a) Modified Moss factor $M_{edge}$ = $E_{edge}n_0^4$ / 95 eV as a function of absorption edge $E_{edge}$ for individual inorganic compounds. Compounds reported in the ICSD database are represented with solid markers whilst compounds with no experimental validation are represented with hollow markers. Compounds marked by stars (*) were synthesized in a crystal structure different than the one from Materials Project which exhibits super-Mossian properties. Compounds with allowed and forbidden transitions at the direct bandgap are represented with round and diamond markers, respectively. (b) Element-based grouping: each marker represents all compounds that contain a given element, plotted as the mean Moss factor versus the mean absorption edge of that element's subset. (c) Structural-prototype grouping: each marker represents all compounds with the same structural prototype. The marker size in (b) and (c) scales linearly with the number of compounds in the corresponding group. Compounds from panel (a) belonging to a group from panel (c) are coloured according to the group colour. The red groups, $Hf_2SeN_2$ (crossed marker) and $Zr_2SeN_2$ (solid marker) correspond to two similar hexagonal prototypes (α and β, respectively) with different space groups (P-3m1 and $P6_3/mmc$, respectively).*

Three other relevant structure prototypes are the $Hf_2SeN_2$, $Zr_2SeN_2$ and ZrSeO, which all host multi-anion compounds based on Group IVB early transition metals (Figure 1c). The ZrSeO prototype contains materials of the (Hf,Zr)(S,Se)O isostructural family. The two other prototypes ($Hf_2SeN_2$ and $Zr_2SeN_2$) obey the $(Hf,Zr)_2(S,Se)N_2$ formula and correspond to two structurally similar but not identical hexagonal polymorphs (α, crossed red marker, and β, solid red marker in Figure 1c), as will be explained later. We have recently confirmed the high refractive index and transparency predicted for one member of this family, $Zr_2SN_2$ [5]. In such structures, the different mechanisms causing low transition dipole moments (incompatible band symmetry, momentum change, poor wavefunction overlap) are present simultaneously. Therefore, the class of $(Hf,Zr)_2(S,Se)N_2$ chalconitrides stands out as a structural prototype promoting high refractive indices and transparency.

For this reason, we further investigate the optical properties of these compounds. In a recent study [5], we discussed the importance of polymorphism in $Zr_2SN_2$. Compounds of the $(Hf,Zr)_2(S,Se)N_2$ family may crystallize in the α or the β phase, two hexagonal structures that only differ in the stacking sequence along the c axis of the $Zr_2N_2$ layers, bonded by S interlayers [26]. We derive more accurate values of the absorption edge $E_{edge}$, the sub-bandgap refractive index $n_0$, and $M_{edge}$ for all the α and β polymorphs of the $(Hf,Zr)_2(S,Se)N_2$ family using hybrid DFT (Figure 2, Table S1). All the α and β polymorphs of this family generally have a large Moss factor. The β phase is consistently more transparent than the α-phase (Figure 2), which generally leads to higher Moss factors for the β compounds. This increase in transparency can be explained by a combination of slightly wider bandgaps, weaker transition dipole moments and stricter symmetry selection rules in β in comparison to α [5]. The substitution of Zr by Hf and Se by S leads to a widening of both the indirect and direct gaps (Table S1, Figure S2-S3).

Across the whole $(Hf,Zr)_2(S,Se)N_2$ family, α-$Hf_2SeN_2$ stands out as an experimentally unreported but chemically plausible compound with a large calculated Moss factor of 2.02 and an HSE absorption edge in the UV, improving on β-$Zr_2SN_2$. Additionally, we find that β-$Hf_2SN_2$, with its larger $E_{edge}$ with minimal compromise on $n_0$, is the compound of choice when a wider transparency window is needed (HSE $E_{edge}$ > 3.7 eV). Notably, β-$Hf_2SN_2$ exhibits the largest offset $\Delta E_{edge}$ between direct bandgap and absorption threshold of all compounds considered, reaching 0.89 eV. If broadband transparency across the visible can be compromised, β-$Zr_2SeN_2$ with a lower $E_{edge}$ at 3.06 eV has the highest Moss factor overall and an ultra-high $n_0$ approaching 2.9.

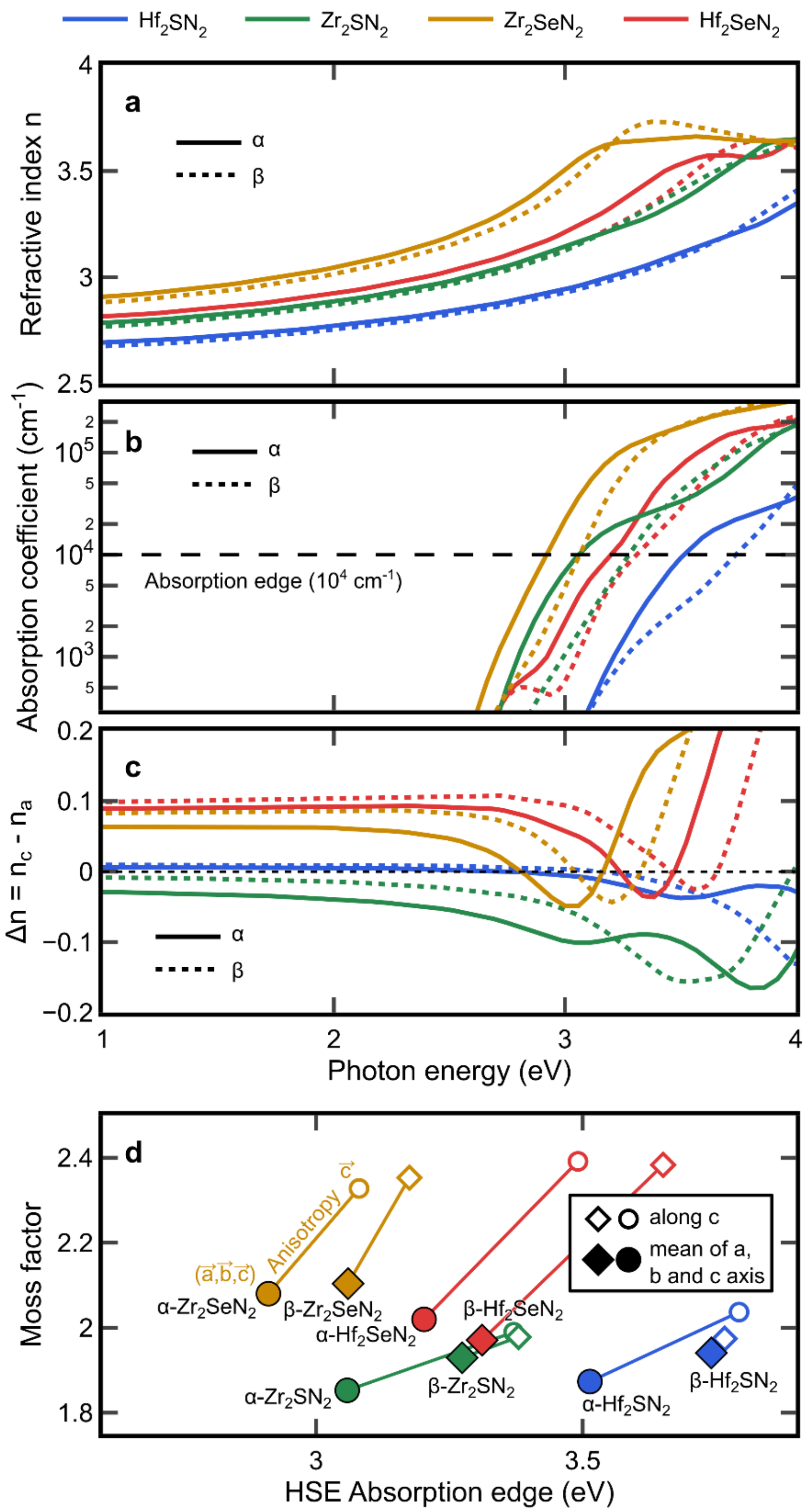

*Figure 2: Calculated HSE06-level optical properties of the (Hf,Zr)$_2$(S,Se)N$_2$ family. Trace colour identifies the chemical formula, while line style distinguishes the α and β polymorphs. Comparison of the (a) refractive index spectra n, (b) absorption coefficient spectra, (c) birefringence $\Delta n = n_c - n_a$ (i.e. the difference between the refractive index $n_c$ and $n_a$ in the c- and a-axis direction), and (d) $M_{edge}$ vs HSE absorption edge for α (circles) and β (diamond) polymorphs. (d) Solid markers correspond to the $M_{edge}$ when averaging over the three crystal directions. Hollow markers correspond to the $M_{edge}$ in the c-axis direction.*

Considering structural anisotropy in the $(Hf,Zr)_2(S,Se)N_2$ family, the c-axis direction systematically displays transparency, with an absorption onset at higher energy for light polarized along c (Table S1), as compared to the symmetry-equivalent a and b axis. As a result, the Moss factor in the c-axis direction is consistently larger (Figure 2d). Birefringence ($\Delta n = n_c - n_a$) is generally larger and positive in Se-containing compounds and weaker and negative in S-containing ones (Figure 2c). Δn in β-$Hf_2SeN_2$ sits at a remarkable and stable + 0.10 at photon energies below 2 eV, because of a larger anisotropy in the β-phase as compared to α-phase. This anisotropy translates into an exceptionally high Moss factor along the c direction, up to 2.38 for β-$Hf_2SeN_2$. For future photonic devices, control of the films' orientation would allow to align the light's polarization axis with the crystal axis with the most favourable $M_{edge}$.

All the effective masses m* for both electrons and holes in $(Hf,Zr)_2(S,Se)N_2$ compounds are low ($0.39 < m^* < 0.46$, Table S1), due to the similar band structures. Low effective masses can be advantageous for indirect-gap high-n materials, because a low joint density of states at the indirect gap will tend to further reduce the already weak indirect absorption in the material. Weak indirect absorption is essential to maintain sufficient optical transparency in indirect gap materials between their fundamental and first allowed direct band gap. This feature can open applications where some absorption can be tolerated (such as metalenses). An additional desirable feature of low effective masses in high n, low k materials is that they may be employed in applications where high conductivity, transparency and refractive index are conjointly needed, as detailed in our recent preprint on $Zr_2SN_2$ [5]. Indeed, combining the low effective masses for both carrier types in $(Hf,Zr)_2(S,Se)N_2$ with previous predictions of ambipolar conductivity [4] and our observation of degenerate conductivity in $Zr_2SN_2$ [5] yields exciting prospects for this family of compounds as high refractive index transparent conductors. This warrants experimental realisation of such compounds as thin films and validation of their dopability and optical properties.

The $E_{edge} = 10^4$ cm$^{-1}$ threshold used throughout this paper is relevant for applications tolerant of moderate residual loss, such as sub-wavelength metasurfaces and localized nanoresonators. In long-propagation-length devices such as integrated waveguides, significantly lower losses are typically required. However, lowering the screening threshold for $E_{edge}$ to the $10^2$-$10^3$ cm$^{-1}$ range already becomes numerically unreliable with the present dataset; absorption spectra from both Materials Project and our in-house HSE06 calculations are Gaussian-broadened to suppress k-point sampling artifacts, so absorption coefficients from the material at these low magnitudes cannot be confidently distinguished from broadening tails. Besides this numerical limitation, at least four physical mechanisms neglected in the present approach could further narrow the true transparency window and reduce the Moss factor figure-of-merit of some of the screened compounds. They are indirect absorption, Urbach tails, excitonic effects, and spin-orbit coupling.

Quantitative treatment of indirect absorption (via electron-phonon coupling) and excitonic effects (GW method with Bethe-Salpeter equation) is computationally too demanding for materials screening purposes and lies outside the scope of this study. Urbach tails from multi-phonon scattering in direct interband transitions are expected to have a magnitude far below the $E_{edge} = 10^4$ cm$^{-1}$ threshold used throughout [27]. Possible implications of these effects on the optical properties of the chalconitrides identified above are discussed in the SI. Spin-orbit coupling (SOC) tends to play a significant role only in certain compounds with heavy elements. To investigate its effect, we have performed a HSE+SOC calculation on α-$Hf_2SeN_2$ as the member of the chalconitride family with the highest average atomic number, and therefore a likely worst-case scenario. The result (Figure S4) is a narrowing of $E_{edge}$ by 0.2 eV, not uncommon for materials with heavy elements, and compensating the average 0.1 eV bandgap underestimation of $(Hf,Zr)(S,Se)_2$ compounds by the HSE06 functional [28]. Based on the discussion above, we regard the HSE-calculated $E_{edge}$ values determined in Figure 2 as an upper bound on the achievable transparency window, most reliable for short-propagation-length applications. Experimental validation will ultimately be required to establish transparency for each candidate compound.

We showed that very large refractive indices can be achieved in compounds from the $(Hf,Zr)_2(S,Se)N_2$ family. Precise control of the α/β polymorphism is important to maximize the potential of these materials. The β phase always has a lower zero-temperature formation energy, but the difference in formation enthalpy $\Delta H_{f,\beta} - \Delta H_{f,\alpha}$ between α and β is minimal (< 18 meV/atom in absolute value, Table S1), hinting that there is no strong thermodynamic drive towards one phase or the other. This is consistent with previous observations in thin film $Zr_2SN_2$ [5], where we evidenced a composite structure composed of epitaxial α and β layers. Experimental validation of thin film samples across the full compound family remains essential, as kinetic effects inherent to thin film growth and rapid thermal annealing, together with entropic contributions at finite temperature, may shift the system's balance between the α and β phases. Importantly, while the β-phase consistently outperforms the α-phase, both polymorphs exhibit large Moss factor and high refractive indices. This means that α/β intergrowth, while non-ideal, will most likely not critically undermine the optical performance of real thin film devices.

## Conclusions

In summary, we introduced a modified Moss rule anchored to the optical absorption edge rather than to the fundamental bandgap, providing a physically meaningful figure of merit for identifying high refractive index, low-loss materials. Screening the Materials Project database for high refractive index compounds uncovers several recurring elements, such as Be, Si, Mg, Zr, Cu, Al, Hf cations and P, S, N anions. There are also recurring structural prototypes, such as (i) pnictide and chalcogenide chalcopyrites, (ii) Zintl pnictides, and (iii) multi-anion chalconitrides and oxichalcogenides of early transition metals. These findings reinforce the motivation to move beyond oxides and single-anion materials in general. Within this landscape, the $(Hf,Zr)_2(S,Se)N_2$ family emerges as a structurally coherent prototype, combining large refractive indices with wide transparency windows. Detailed hybrid DFT analysis of the full α/β polymorph space reveals that α-$Hf_2SeN_2$ achieves a sub-bandgap refractive index of 2.78 and an absorption edge in the UV. Notably, β-$Hf_2SeN_2$ displays birefringence of + 0.1 between the c-axis and a-axis, yielding even larger Moss factors if crystal orientation can be controlled. β-$Zr_2SeN_2$ with a lower $E_{edge}$ at 3.06 eV displayed an ultra-high sub-bandgap refractive index of 2.84. β-$Hf_2SN_2$ offers the broadest transparency window across the family. These findings establish $(Hf,Zr)_2(S,Se)N_2$ chalconitrides as a compelling material family for next-generation photonic applications and motivate experimental thin film synthesis efforts aimed at controlling polymorphism and crystal orientation through careful tuning of thin film growth conditions and thermal processing.

## Acknowledgements

We are grateful to Søren Raza for insightful discussions.

## Funding Sources

This work was funded by the European Union (ERC, IDOL, 101040153). Views and opinions expressed are, however, those of the authors only and do not necessarily reflect those of the European Union or the European Research Council. Neither the European Union nor the granting authority can be held responsible for them.

This work was also supported by a research grant (42140) from VILLUM FONDEN.

## Associated content

**Preprint**

A previous version of this manuscript has been deposited on the arXiv server:

Bertin, E.; Tadge, F.; Crovetto, A. A Modified Moss Rule Highlights Underexplored Classes of High Refractive Index Materials. 2026, 2607.07086. arXiv. https://doi.org/10.48550/arXiv.2607.07086 (accessed October 19, 2026).

## Supporting Information

Modified Moss factor of binary inorganic compounds; HSE06 band structures, structural, electronic and optical properties of $(Hf,Zr)_2(S,Se)N_2$ polymorphs; comparison with benchmark high-index materials; discussion of indirect absorption, excitonic, spin-orbit coupling, and Urbach tail effects; column definitions of database-SI.csv, a dataset of refractive indices, band gaps, absorption edges, Moss factors, and stability and structural properties of many inorganic compounds.

# A modified Moss rule highlights underexplored classes of high refractive index materials

Eugène Bertin*, Finja Tadge, Andrea Crovetto*

*National Centre for Nano Fabrication and Characterization (DTU Nanolab), Technical University of Denmark*

**corresponding authors*

**For Table of Contents Use Only**

## Table of Content Graphic

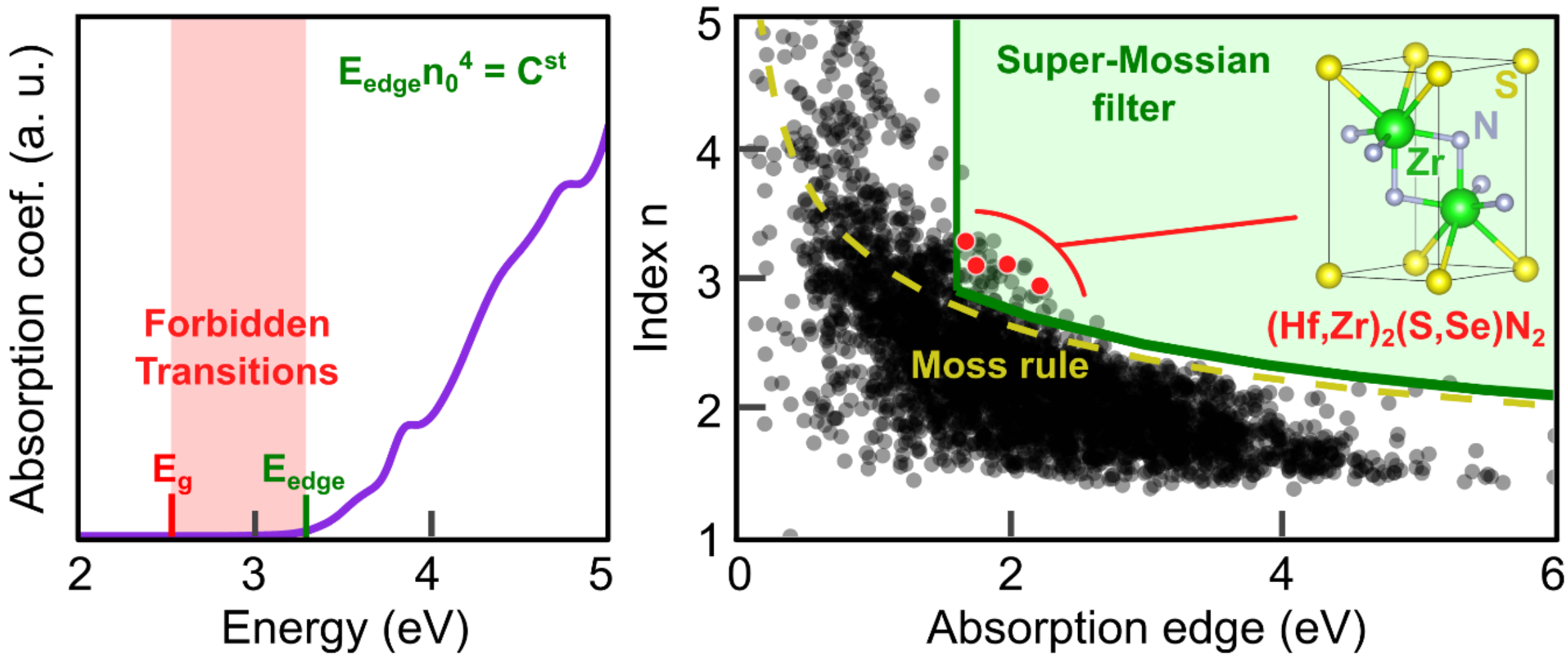

This ToC graphic presents four key concepts (i-iv) and results of the work:

- The left panel illustrates (i) how the onset of the optical absorption $E_{edge}$ can occur at higher photon energies than the bandgap energy $E_G$, if electronic transitions at the bandgap edges are momentum- or symmetry-forbidden or if wavefunction overlap between initial and final states is poor. (ii) It also presents a revised Moss rule in which the characteristic energy is the absorption edge $E_{edge}$ instead of the bandgap, yielding a figure-of-merit highlighting materials with high n and high absorption edge (super-Mossian materials)
- The right panel illustrates (iii) how compounds perform within this new framework. (iv) In particular, a family of $(Hf,Zr)_2(S,Se)N_2$ chalconitride compounds, stand out as potential super-Mossian materials.

# SUPPORTING INFORMATION:

# A modified Moss rule highlights underexplored classes of high refractive index materials

Eugène Bertin, Finja Tadge, Andrea Crovetto

*National Centre for Nano Fabrication and Characterization (DTU Nanolab), Technical University of Denmark*

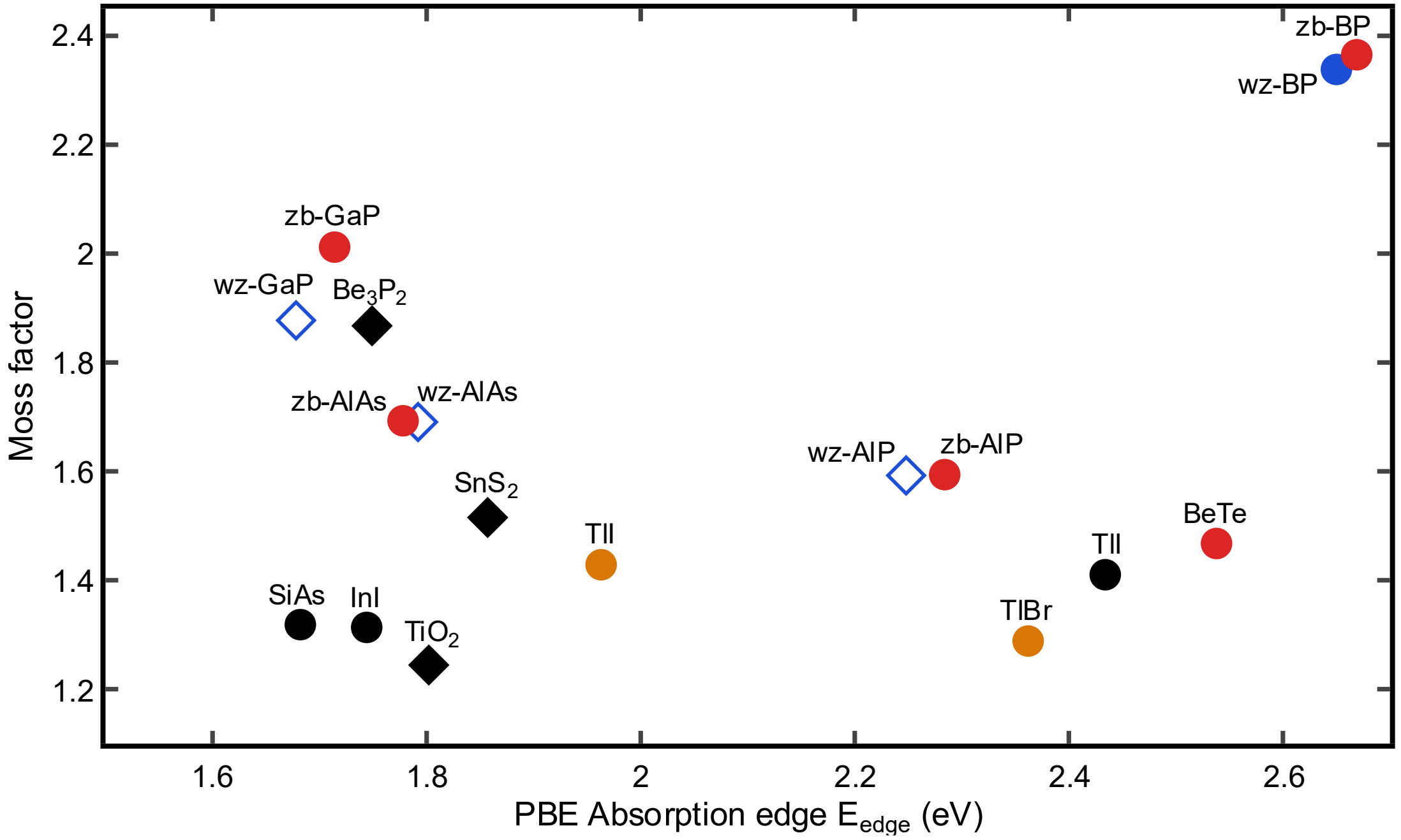

*Figure S1: Modified Moss factor $M_{edge} = E_{edge}n_0^4 / 95$ eV as a function of PBE absorption edge $E_{edge}$ for binary inorganic compounds. Compounds reported in the ICSD database are represented with solid markers whilst compounds with no experimental validation are represented with hollow markers. Compounds with allowed and forbidden transitions at the direct bandgap are represented with round and diamond markers, respectively. zb and wz labels correspond to zinc-blende (cubic) and wurtzite (hexagonal) prototypes, respectively.*

*Table S1: Summary of the optical and thermodynamic properties for $(Hf,Zr)_2(S,Se)N_2$ polymorphs. For each material and polymorph, the table reports the lattice parameters a and c, the effective masses m* for electron $m^*_e$ and holes $m^*_h$, given along particular reciprocal lattice vectors and as averaged (harmonic mean) along all reciprocal lattice vectors. The sub-bandgap refractive index $n_0$, the Penn gap $E_P$, the indirect band gap $E^i_g$ (from K to Γ special points) and direct bandgap $E^d_g$ (at the K special point), the absorption edge $E_{edge}$ (defined as the energy where the absorption coefficient surpasses $10^4$ cm$^{-1}$), the energy difference $\Delta E_{edge}$ between the absorption edge $E_{edge}$ and the direct band gap $E^d_g$, the modified Moss factor $M_{edge} = E_{edge}n_0^4/95eV$, and the formation energy difference $\Delta H_f$ between β and α. The directional $n_0$, $E_{edge}$ and $M_{edge}$ in the a- and c-axis direction of the crystal are also given and marked by subscripts.*

| Compound | a(Å) | c(Å) | $m^*_e$ from K to Γ/H | avg $m^*_e$ | $m^*_h$ from Γ to A/M/K | avg $m^*_h$ |
|---|---|---|---|---|---|---|
| α-$Hf_2SN_2$ | 3.553 | 6.345 | 0.34/0.69 | 0.46 | 0.30/0.49/0.49 | 0.41 |
| β-$Hf_2SN_2$ | 3.557 | 12.627 | 0.34/0.71 | 0.46 | 0.31/0.51/0.51 | 0.42 |
| α-$Hf_2SeN_2$ | 3.583 | 6.569 | 0.33/0.62 | 0.43 | 0.25/0.54/0.55 | 0.39 |
| β-$Hf_2SeN_2$ | 3.590 | 13.042 | 0.34/0.64 | 0.44 | 0.25/0.54/0.54 | 0.39 |
| α-$Zr_2SN_2$ | 3.594 | 6.371 | 0.33/0.74 | 0.45 | 0.36/0.48/0.48 | 0.43 |
| β-$Zr_2SN_2$ | 3.596 | 12.719 | 0.33/0.75 | 0.46 | 0.36/0.52/0.51 | 0.45 |
| α-$Zr_2SeN_2$ | 3.624 | 6.600 | 0.32/0.67 | 0.43 | 0.29/0.53/0.54 | 0.42 |
| β-$Zr_2SeN_2$ | 3.629 | 13.128 | 0.32/0.68 | 0.44 | 0.30/0.54/0.54 | 0.43 |

| Compound | $n_0$ | $n_{0,a}$ | $n_{0,c}$ | $E_P$ (eV) | $E_g^i$ (eV) | $E_g^d$ (eV) | $E_{edge}$ (eV) | $\Delta E_{edge}$ (eV) | $E_{edge,a}$ (eV) | $E_{edge,c}$ (eV) | $M_{edge}$ | $M_{edge,a}$ | $M_{edge,c}$ | $\Delta H_{f,\beta} - \Delta H_{f,\alpha}$ (meV/atom) |
|---|---|---|---|---|---|---|---|---|---|---|---|---|---|---|
| α-$Hf_2SN_2$ | 2.67 | 2.67 | 2.67 | 6.09 | 1.72 | 2.96 | 3.51 | 0.55 | 3.47 | 3.79 | 1.87 | 1.84 | 2.04 | -17.71 |
| β-$Hf_2SN_2$ | 2.65 | 2.65 | 2.66 | 6.18 | 1.73 | 2.86 | 3.74 | 0.89 | 3.73 | 3.77 | 1.94 | 1.93 | 1.97 | |
| α-$Hf_2SeN_2$ | 2.78 | 2.75 | 2.84 | 5.85 | 1.39 | 2.81 | 3.20 | 0.39 | 3.17 | 3.49 | 2.02 | 1.92 | 2.39 | -15.83 |
| β-$Hf_2SeN_2$ | 2.74 | 2.71 | 2.81 | 6.01 | 1.41 | 2.78 | 3.31 | 0.53 | 3.28 | 3.65 | 1.97 | 1.86 | 2.38 | |
| α-$Zr_2SN_2$ | 2.76 | 2.76 | 2.74 | 6.01 | 1.43 | 2.62 | 3.06 | 0.44 | 3.00 | 3.37 | 1.85 | 1.84 | 1.99 | -13.62 |
| β-$Zr_2SN_2$ | 2.74 | 2.74 | 2.73 | 6.09 | 1.43 | 2.63 | 3.27 | 0.64 | 3.26 | 3.38 | 1.93 | 1.92 | 1.98 | |
| α-$Zr_2SeN_2$ | 2.87 | 2.85 | 2.91 | 5.75 | 1.13 | 2.50 | 2.92 | 0.42 | 2.89 | 3.08 | 2.08 | 2.00 | 2.33 | -15.04 |
| β-$Zr_2SeN_2$ | 2.84 | 2.82 | 2.90 | 5.91 | 1.15 | 2.50 | 3.06 | 0.56 | 3.04 | 3.18 | 2.10 | 2.01 | 2.35 | |

**Comparison between high-index transparent materials**

We collected the figure of merit for $TiO_2$, SiC, GaN, $Si_3N_4$, as well several chalconitrides, in the following Table S2 by querying the absorption coefficients calculated by Woods-Robinson et al. [4] at the PBE-DFT level (available on MPContribs) and calculating the Moss factor $M_{edge}$ from absorption edge energy $E_{edge}$ (@ $10^4$ $cm^{-1}$). This is same procedure as the one conducted in the main article. We find an inconsistency in the PBE absorption coefficient available for SiC on MPContribs: its absorption edge $E_{edge}$ lies at a lower photon energy than its direct bandgap, which is unphysical in the absence of an indirect absorption calculation. Therefore, we recalculate the 3C-SiC dielectric constants at the PBE level. In our calculation, $E_{edge}$ in SiC lies above the direct bandgap as expected and we recalculate a $M_{edge}$ of 2.43 based our PBE calculation, instead of the original 1.81 based on MPContribs. Based on the modified Moss metric $M_{edge}$ introduced in the main article (Table S2), chalconitrides seem to outperform rutile-$TiO_2$, as well as other established dielectrics.

*Table S2: Sub-bandgap refractive index $n_0$, absorption edge energy $E_{edge}$ (@ $10^4$ $cm^{-1}$) and corresponding Moss factor $M_{edge}$ for different benchmark materials. With the exception of SiC, the data is queried from Materials Project MPContribution from Woods-Robinson [4]. *SiC optical properties were recalculated at the PBE level due to the inconsistency in the MPContribution.*

| Compound | mp-id | $n_0$ | $E_{edge}$ (@ $10^4$ $cm^{-1}$) (eV) | $M_{edge}$ |
|---|---|---|---|---|
| $TiO_2$ (rutile) | mp-2657 | 2.85 | 1.80 | 1.24 |
| SiC (3C)* | mp-8062 | 2.67 | 4.53 | 2.43 |
| GaN (wurtzite) | mp-804 | 2.44 | 1.81 | 0.68 |
| $Si_3N_4$ (hexagonal) | mp-988 | 2.07 | 4.57 | 0.87 |
| α-$Hf_2SN_2$ | mp-1017567 | 2.94 | 2.21 | 1.74 |
| α-$Hf_2SeN_2$ | mp-1029330 | 3.10 | 1.96 | 1.91 |
| β-$Zr_2SN_2$ | mp-11583 | 3.09 | 1.87 | 1.79 |
| β-$Zr_2SeN_2$ | mp-1079726 | 3.27 | 1.66 | 2.01 |

**Electronic band structures for α and β polymorphs of the (Hf,Zr)$_2$(S,Se)$N_2$ system**

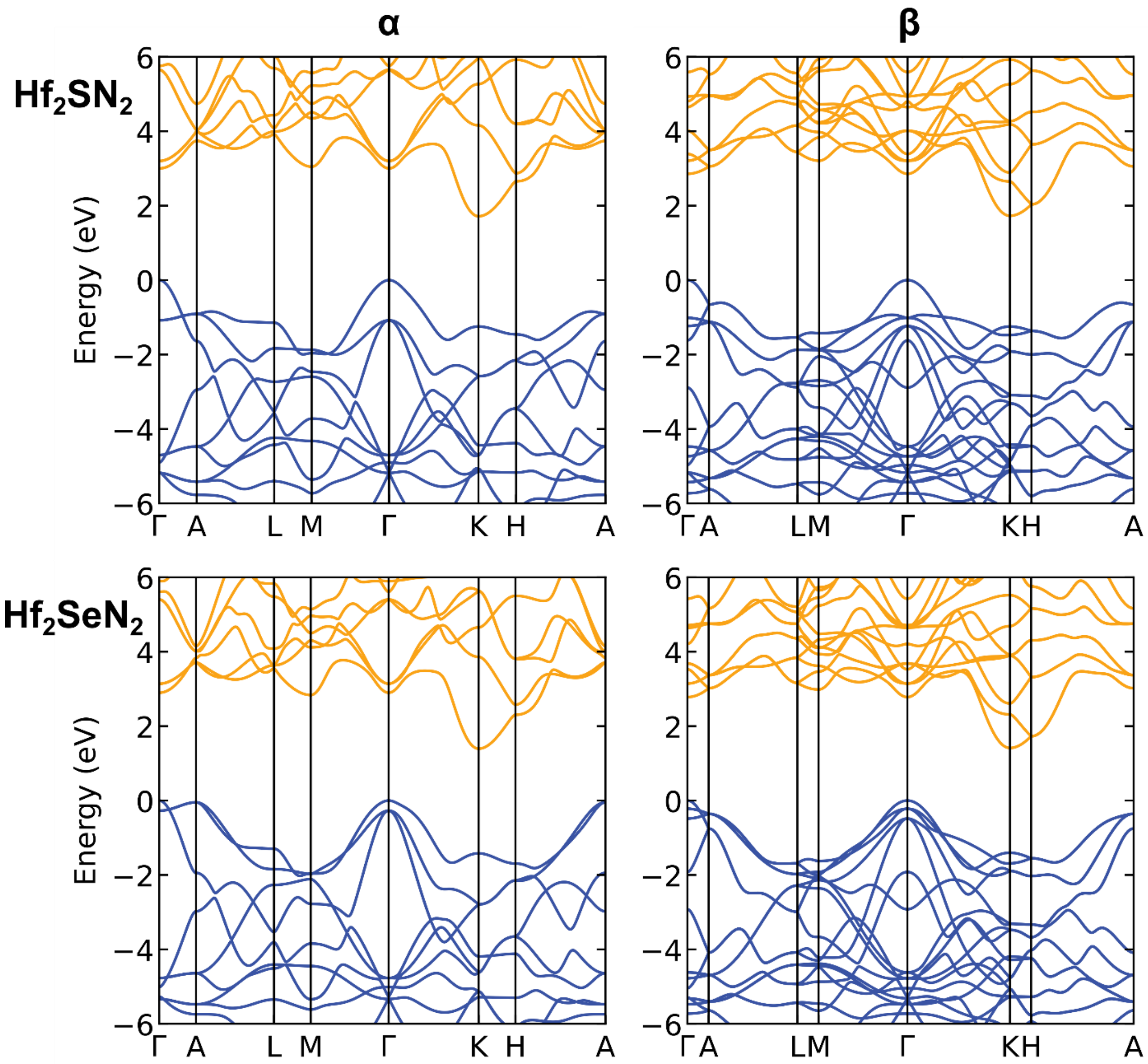

*Figure S2: Calculated HSE06-level electronic band structures of the $Hf_2(S,Se)N_2$ family in the α and β structures.*

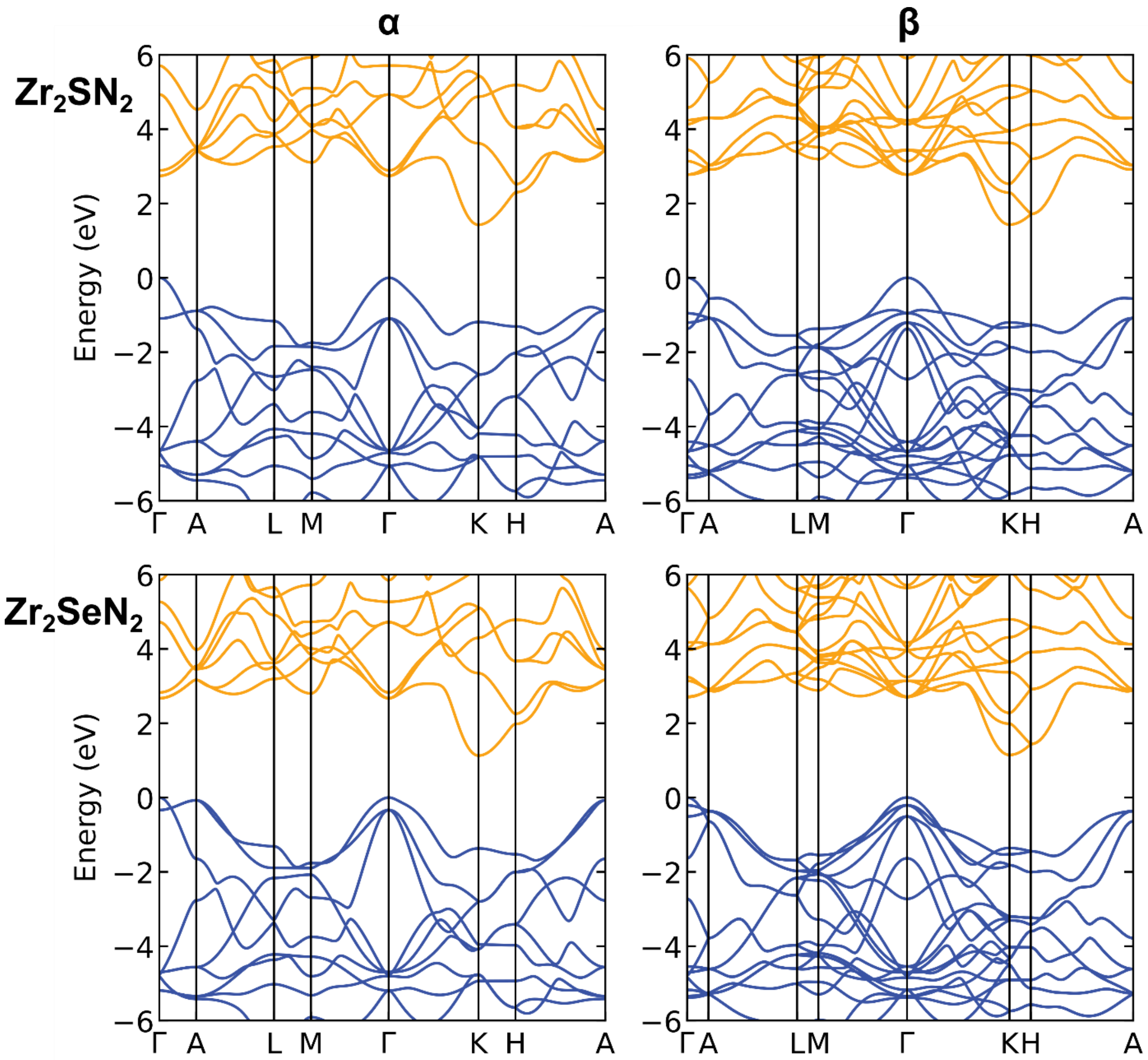

*Figure S3: Calculated HSE06-level electronic band structures of the $Zr_2(S,Se)N_2$ family in the α and β structures.*

**Discussion of the possible effects of indirect absorption, Urbach tails, excitonic absorption, and spin-orbit coupling**

At least four physical mechanisms neglected in the present approach could further narrow the true transparency window and reduce the Moss factor figure-of-merit of some of the screened compounds. They are indirect absorption, excitonic effects, and spin-orbit coupling and Urbach tails.

The first mechanism is phonon-assisted indirect absorption, especially given the modest indirect gaps of the $(Hf,Zr)_2(S,Se)N_2$ family (1.13–1.73 eV, Table S1). For the structurally related compounds $ZrS_2$ and $HfS_2$, reported indirect absorption coefficients reach hundreds of $cm^{-1}$. However, the joint density of states at the indirect edge is likely low in the $(Hf,Zr)_2(S,Se)N_2$ family, owing to their low and isotropic effective masses (Table S1). This feature may cause weaker indirect absorption than in $ZrS_2$ and $HfS_2$.

The second mechanism is excitonic effects, expected to red-shift the true absorption onset with respect to the calculated $E_{edge}$. To the best of our knowledge, the literature on excitonic effects in chalconitrides and related compounds is scarce. Bulk $ZrS_2$ and $HfS_2$ have been experimentally characterized [29] and although features in the absorption spectra are detected, no clear excitonic resonance could be resolved. Hf and Zr chalconitrides are reported to have rather low exciton binding energy [30]. Heuristically, the large dielectric constant and low effective masses for both holes and electrons in $(Hf,Zr)_2(S,Se)N_2$ points to low exciton binding energy for all compounds.

The third mechanism is spin-orbit coupling (SOC) that can lead to a significant alteration of transition dipole moments due to lifting band degeneracies and splitting of bands, especially in compounds containing heavy transition metals (like Hf, Zr) and chalcogenides (Se). We performed HSE06 level calculations of the electronic band structure and the absorption coefficient including SOC for the $\alpha$-$Hf_2SeN_2$ phase to estimate the influence of SOC on the electronic and optical properties of the $(Hf,Zr)_2(S,Se)N_2$ family. This phase was chosen as an extreme case representative of this family, as it comprises both the heavy transition metal Hf and Se.

Figure S4 shows the resulting band structure as well as absorption coefficient upon inclusion of SOC as compared to the results obtained when SOC is not included. Taking SOC into consideration leads to an expected splitting and energy shifts of several bands in the $\alpha$-$Hf_2SeN_2$ band structure; however, the overall electronic structure is only moderately affected. The magnitude of the indirect band gap decreases by only 0.13 eV, while the direct band gap remains essentially unchanged, deviating by only 0.02 eV. Assuming that indirect absorption is expected to be weak enough to be tolerated, this decrease in the indirect band gap energy is not expected to prematurely close the transparent region.

The absorption coefficient calculated with inclusion of SOC shows a similar shape to the absorption coefficient without SOC, however, it is shifted to slightly lower energies by around 0.2 eV. This leads to a shift in the absorption edge ($E_{edge}$) of 0.2 eV from 3.2 eV to 3.0 eV for the $\alpha$-$Hf_2SeN_2$ phase. The magnitude of this shift is expected upon the inclusion of SOC and is approximately of the same magnitude as the difference in $E_{edge}$ between the α- and β-phases within members of the $(Hf,Zr)_2(S,Se)N_2$ family.

These results indicate that the quantitative effect of the inclusion of SOC on the calculated band gaps and absorption coefficient is negligible for α-$Hf_2SeN_2$. We thus consider our principal conclusions regarding the optical properties of the chalconitride material family to be robust with respect to SOC.

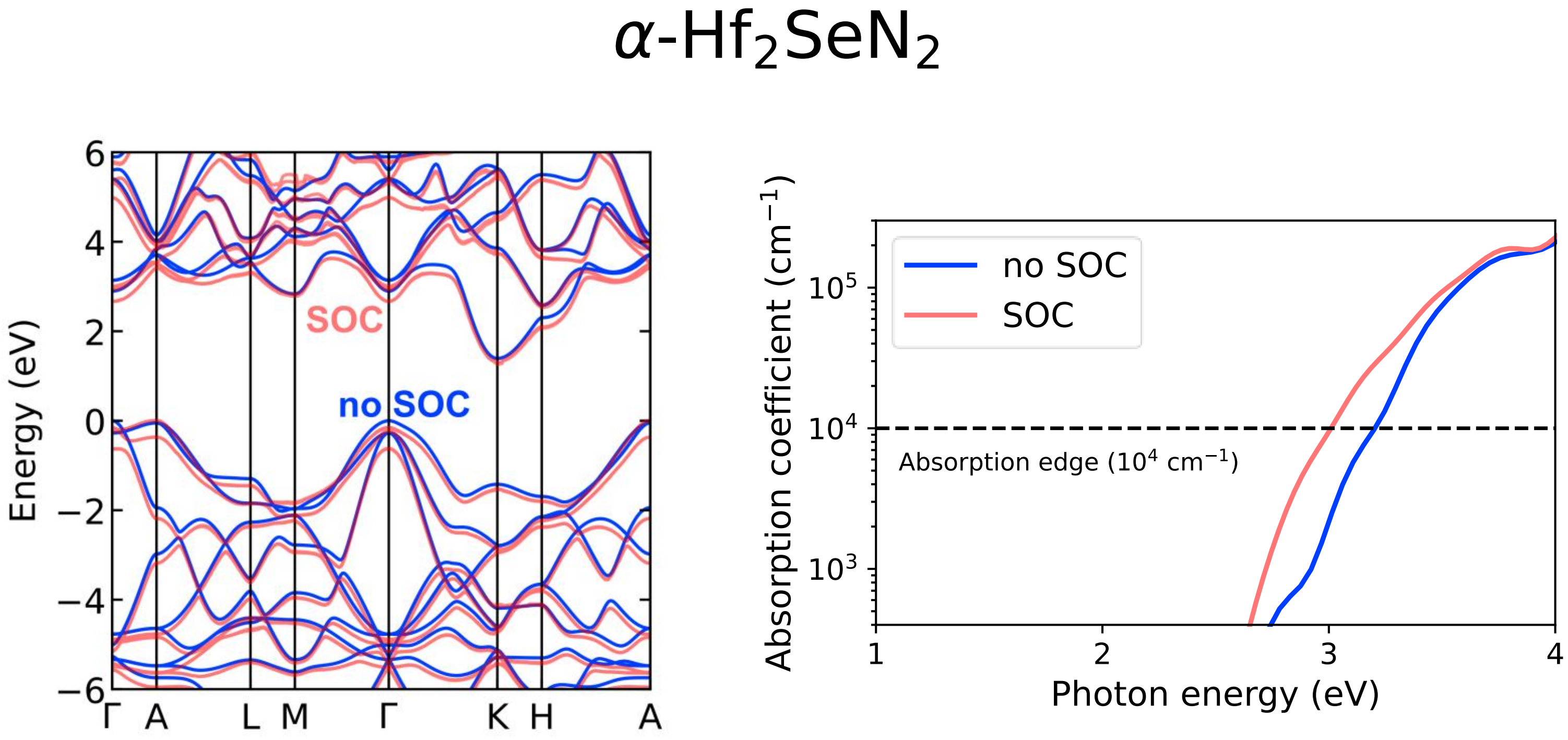

*Figure S4: Calculated HSE06 level band structure and absorption coefficient for the α-$Hf_2SeN_2$ phase considering the inclusion of spin-orbit coupling (SOC, red line) compared to calculations performed without spin-orbit coupling (no SOC, blue line).*

Another effect is direct transition Urbach absorption tails. To gauge the magnitude of the Urbach absorption tail, we use the theory developed by Khurgin [27], where a theory of Urbach tails in direct optical transitions resulting from multi-phonon scattering is presented. Calculations of phonon band structures and Raman spectra for α- and β-$Zr_2SN_2$ are available in the Computational Raman Database [31]. In a recent preprint [5] we show that the experimental Raman spectra of $Zr_2SN_2$ can be indexed accurately using the computed non-polar Raman-active modes from the database. This validates the calculated phonons energies in the database.

Among the polar IR-active phonons, one in-plane phonon stands out as a strong scatterer. The $E_u$ and $E_{1u}$ modes undergo the largest LO/TO splitting at the Γ point, from 498 to 684 $cm^{-1}$, and from 508 to 687 $cm^{-1}$ for α- and β-$Zr_2SN_2$, respectively. This splitting results in a LO mode with the highest frequency $\Omega_{LO}$ at 684 and 687 $cm^{-1}$ in α- and β- $Zr_2SN_2$ respectively, which corresponds to an energy of $\hbar\Omega_{LO}$ = 85 meV. Therefore, this mode is expected to be the main contributor to the Urbach tail. In the two phonon-limit, the Urbach tail would extend $2\hbar\Omega_{LO}$ = 170 meV below the direct bandgap.

To gauge the magnitude of the Urbach tail, we calculate the Fröhlich interaction $\frac{1}{\epsilon'}$ as

$$\frac{1}{\epsilon'} = \frac{1}{\epsilon_\infty} - \frac{1}{\epsilon_0},$$

with $\epsilon_\infty$ the high frequency dielectric constant, and $\epsilon_0$ the static dielectric constant. In the single-oscillator approximation using the phonon mode identified earlier, can be estimated using the Lyddane-Sachs-Teller equation,

$$\epsilon_0 = \epsilon_\infty \left(\frac{\Omega_{\mathrm{LO}}}{\Omega_{\mathrm{TO}}}\right)^2 ,$$

with $\Omega_{\mathrm{LO}}$ and $\Omega_{\mathrm{TO}}$ the frequency of the longitudinal and transverse optical phonon branches respectively. This yields an estimated Fröhlich interaction of 0.048 and 0.047 in α- and β-$Zr_2SN_2$, respectively.

The effective carrier masses in $Zr_2SN_2$ are rather low ($m_e^* \approx 0.40$, at K where the direct transition occurs, for both α and β phases), which limits the strength of the Fröhlich coupling. The dimensionless Fröhlich coupling constant $\alpha_{\mathrm{F}}$ scales as

$$\alpha_{\mathrm{F}} \propto \left(\frac{1}{\epsilon'}\right) \sqrt{\frac{\mathrm{m}^*}{2\hbar\Omega_{\mathrm{LO}}}},$$

with m* the carrier effective mass (here 0.40, using electron's effective mass; see Materials Project entries mp-11583 and mp-553875), and $\hbar$ the reduced Planck constant. In the single-oscillator approximation, we estimate a $\alpha_F$ between 0.1 and 1, which corresponds to an intermediate coupling. This is similar to GaN and CdS [27] and a broad but weak phonon-assisted tail is expected. Therefore, Urbach tails from multi-phonon scattering in direct interband transitions are expected to have a magnitude far below the $E_{edge} = 10^4$ $cm^{-1}$ threshold used throughout.

**database-SI.csv: Woods-Robinson and Materials Project Dataset Variables**

The database-SI.csv file available for download as Supplementary Information combines Woods-Robinson *ForbiddenTransitions* descriptors with Materials Project dielectric/summary data and derived quantities. Blank values mean the source did not provide a value, the corresponding MP ID could not be enriched, or the calculation was not applicable.

The Woods-Robinson fields originate from the MPContribs ForbiddenTransitions project, associated with Woods-Robinson et al., *Matter* (2023), DOI: 10.1016/j.matt.2023.06.043 [4].

The meaning of the different column header names can be found in the following table

*Table S3: Definitions and Sources of Variables in the Woods-Robinson–Materials Project Combined Dataset*

| **Column** | **Units / type** | **Source** | **Meaning / calculation** |
|---|---|---|---|
| MP_id | string | MPContribs | Materials Project material identifier used to join all datasets. IDs are normalized to lowercase in this script. |
| Formula | string | MP dielectric document; fallback MPContribs | Reduced chemical formula (formula_pretty). |

| Column | Units / type | Source | Meaning / calculation |
|---|---|---|---|
| n | dimensionless | MP dielectric document | Isotropic high-frequency refractive index. The script uses an explicit refractive-index field when present, otherwise $n_\infty = \sqrt{\langle \epsilon_\infty \rangle}$. |
| wr_Eg_fund | eV | MPContribs | Fundamental electronic band gap, $E_g^{fund}$, including indirect transitions. |
| wr_Ed_G | eV | MPContribs | Fundamental direct gap, $E_g^d$: lowest vertical transition, which can be optically forbidden. |
| wr_Eda_G | eV | MPContribs | Direct allowed gap, $E_g^{da}$: lowest direct transition with appreciable dipole matrix elements. In the WR workflow it includes the PBE-to-HSE gap correction. |
| wr_Eedge | eV | MPContribs | Absorption-edge energy, $E_{edge}$: energy at which the independent-particle-approximation absorption coefficient first exceeds $10^4\ \mathrm{cm}^{-1}$. |
| wr_delta_d | eV | MPContribs | Forbidden-transition energy difference, $\Delta^d = E_g^{da} - E_g^d$. A positive value indicates a direct band-edge transition that is dipole-forbidden. |
| wr_a_vis | $cm^{-1}$ | MPContribs | Visible-range average absorption coefficient, $\bar{\alpha}_{vis}$, computed from the gap-corrected absorption spectrum. |

| Column | Units / type | Source | Meaning / calculation |
|---|---|---|---|
| Moss_factor_WR | dimensionless | Script | Modified Moss factor $M = n_{analysis}^4 E_{edge}/95$, with $E_{edge}$ in eV. |
| epsilon_infinity_x | dimensionless | MP dielectric document | Electronic dielectric-tensor diagonal component $\epsilon_{\infty,xx}$. |
| epsilon_infinity_y | dimensionless | MP dielectric document | Electronic dielectric-tensor diagonal component $\epsilon_{\infty,yy}$. |
| epsilon_infinity_z | dimensionless | MP dielectric document | Electronic dielectric-tensor diagonal component $\epsilon_{\infty,zz}$. |
| n_x | dimensionless | Script from MP dielectric | Principal refractive index $\sqrt{\epsilon_{\infty,xx}}$ when the component is positive. |
| n_y | dimensionless | Script from MP dielectric | Principal refractive index $\sqrt{\epsilon_{\infty,yy}}$ when the component is positive. |
| n_z | dimensionless | Script from MP dielectric | Principal refractive index $\sqrt{\epsilon_{\infty,zz}}$ when the component is positive. |
| n_anisotropic_mean | dimensionless | Script | Arithmetic mean of the available n_x, n_y, and n_z values. |
| n_anisotropic_max | dimensionless | Script | Largest available principal refractive index, max(n_x, n_y, n_z). |

| Column | Units / type | Source | Meaning / calculation |
|---|---|---|---|
| is_anisotropic_refractive_index | boolean / missing | Script | True when all three principal indices are available and not equal within numerical tolerance; False for isotropic tensors; missing when the tensor is incomplete. |
| Moss_factor_WR_anisotropic | dimensionless | Script | Direction-optimistic Moss factor $n_{max}^4 E_{edge}/95$, using n_anisotropic_max. |
| is_experimental | boolean / missing | MP summary document | True when MP reports theoretical=False; False when theoretical=True. It means the MP entry is experimentally reported, not that this phase was independently reproduced here. |
| exp_label | categorical | Script from is_experimental | Human-readable status: Experimental, Theoretical, or Unknown. |
| epsilon_inf | dimensionless | Script | Effective dielectric constant used for filtering, $\epsilon_\infty = n_{analysis}^2$. It is not necessarily the same as any individual tensor component. |
| spacegroup_symbol | Hermann-Mauguin string | MP summary document | Space-group symbol for the MP structure. |
| spacegroup_number | integer | MP summary document | International space-group number (1-230). |

| Column | Units / type | Source | Meaning / calculation |
| --- | --- | --- | --- |
| crystal_system | categorical | MP summary document | Crystal system reported by MP, such as cubic, tetragonal, or monoclinic. |
| energy_above_hull | eV/atom | MP summary document | Thermodynamic energy above the convex hull. 0 indicates a hull-stable MP entry; positive values indicate metastability within the MP chemical-potential model. |
| is_stable | boolean / missing | MP summary document | Materials Project stability flag, normally true when the entry lies on the convex hull. |
| formation_energy_per_atom | eV/atom | MP summary document | Formation energy per atom relative to Materials Project reference elemental phases. |
| nsites | integer | MP summary document | Number of atomic sites in the MP structure cell represented by the summary document. |
| density | g/cm$^3$ | MP summary document | Calculated mass density of the MP structure. |
| volume | Å$^3$ | MP summary document | Calculated unit-cell volume of the MP structure. |